\documentclass[aps,pra,twocolumn,showpacs,superscriptaddress]{revtex4}%
\usepackage{amsfonts}
\usepackage{amsmath}
\usepackage{amssymb}
\usepackage{graphicx}%
\begin{document}
\title{Reduced fidelity susceptibility in the
one-dimensional transverse field Ising model}
\author{Jian Ma}
\affiliation{Zhejiang Institute of Modern Physics, Department of Physics, Zhejiang
University, HangZhou 310027, P.R. China.}
\author{Lei Xu}
\affiliation{Zhejiang Institute of Modern Physics, Department of Physics, Zhejiang
University, HangZhou 310027, P.R. China.}
\author{Xiaoguang Wang}
\email{xgwang@zimp.zju.edu.cn}
\affiliation{Zhejiang Institute of Modern Physics, Department of Physics, Zhejiang
University, HangZhou 310027, P.R. China.}
\date{\today }

\begin{abstract}
We study critical behaviors of the reduced fidelity susceptibility
for two neighboring sites in the one-dimensional transverse field
Ising model. It is found that the divergent behaviors of the
susceptibility take the form of square of logarithm, in contrast
with the global ground-state fidelity susceptibility which is power
divergence. In order to perform a scaling analysis, we take the
square root of the susceptibility and determine the scaling exponent
analytically and the result is further confirmed by numerical
calculations.

\end{abstract}

\maketitle

\section{Introduction}

Within the last few years, some concepts and tools in the field of
quantum information science \cite{NielsenQIQC}, such as entanglement
and fidelity, have been introduced in the study of quantum phase
transitions (QPTs) \cite{SachdevQPTs}, which are described
traditionally in terms of order parameters and symmetry breaking.
However, without \textit{a} \textit{priori} knowledge of the order
parameter, when a system undergoes QPT, its ground state will change
dramatically. Since the definition of fidelity is the overlap
between two quantum states, it's natural to introduce it in the
field of QPTs. The most studied fidelity is the global ground state
fidelity
\cite{quan:140604,zanardi:031123,buonsante:110601,Zanardi:L02002,
cozzini:014439,venuti:095701,Paolo:PRL100603,you:022101,Chen:arxiv/0801.0020,zhou:cond-mat/0701608,zhou:0704.2940,
gu:0706.2495,Chen:arxiv/0706.0072,
Chen:arxiv/0708.3178,Kwok:arxiv/0710.2581v1,yang:180403,Zhao:arxiv/0803.0814,
Yang:arxiv/0803.1292,zhou:0704.2945}, which is an overlap between
two ground states with a slight change of the parameters. Therefore,
to avoid the arbitrariness of the small change in numerical
computation, Zanardi et al. introduced the Riemannian metric tensor
\cite{venuti:095701,Paolo:PRL100603}, while You et al. suggested the
fidelity susceptibility \cite{you:022101}, which is an effective
tool to study the critical properties.

So far, in most works, people have concerned with the global ground
state fidelity. We note that as far as in
Ref~\cite{Znidaric:JPhysA2463} the authors have introduced a reduced
fidelity which is defined as the trace of the product of two
matrices. This definition is not rigorous in quantum-information
theory, and they require one of the density matrices to be a pure
state to agree with the rigorous mixed state fidelity
\cite{Uhlmann:rep-math-phys}. While in this paper, the reduced
fidelity we used is defined rigorously in quantum-information theory
and is studied in
\cite{zhou:0704.2945,gorin:quant-ph/0607050,Paunkovic:arxiv/0708.3949,JianMa:arxiv/0805.4062v1,S-J-Gu:quant-ph0805}.
It has been shown that the reduced fidelity susceptibility (RFS) is
an efficient indicator of QPTs.

In this paper, we study the scaling and critical behaviors of the
RFS of two neighboring sites in the famous one-dimensional (1D)
transverse field Ising model (TFIM). We note that the quantum
critical properties of some other quantities like concurrence
~\cite{Osterloh:Nature416} and von Neumann entropy
~\cite{S-J-Gu:PRA032308}, and the global ground state fidelity
susceptibility~\cite{Chen:arxiv/0801.0020} have been studied in this
model. However, the scaling and critical behaviors of them are quite
different from the RFS for two neighboring sites . It's found that
the scaling and critical behaviors of the RFS take the form of
square of logarithm as $(\ln N)^{2}$ for finite-size system and
$(\ln |\lambda-1|)^{2}$ ($N$ is the total number of spins and
external parameter $\lambda=1$ is the critical point) in the
thermodynamic limit, in contrast with the global fidelity
susceptibility \cite{Chen:arxiv/0801.0020}, which takes the form of
$N$ and $|1-\lambda|^{-1}$, respectively. The different scaling and
critical behaviors lead to a different finite-size scaling analysis,
and for RFS we have to take its square root which is logarithmic
divergent to carry out the scaling analysis.

The paper is organized as follows. In Sec. II, we briefly review some concepts
in fidelity and its susceptibility, and then we introduce a general formula of
the RFS derived in \cite{JianMa:arxiv/0805.4062v1}. In Sec. III, the scaling
and critical behaviors are obtained analytically for both finite-size and
thermodynamic limit, respectively. Then, using the square root of the RFS, we
perform a finite-size scaling analysis. The scaling exponent is determined
analytically, and is confirmed numerically.

\section{Reduced Fidelity Susceptibility}

Firstly, let us review some concepts in fidelity susceptibility. The
Hamiltonian of a quantum system undergoing QPTs can be written as%
\begin{equation}
H\left(  h\right)  =H_{0}+hH_{I},
\end{equation}
where $H_{I}$ is supposed to be the driving term with a control
parameter $h$. The fidelity between two RDMs of the ground state
$\rho\equiv\rho\left( h\right) $ and $\tilde{\rho}\equiv\rho\left(
h+\delta\right)  $ is defined as \cite{Uhlmann:rep-math-phys}
\begin{equation}
F\left(  h,\delta\right)  =\text{tr}\sqrt{\rho^{1/2}\tilde{\rho}\rho^{1/2}}.
\end{equation}
The fidelity susceptibility is given by
\cite{zanardi:031123,you:022101}
\begin{equation}
\chi=\lim_{\delta\rightarrow0}\frac{-2\ln F}{\delta^{2}}, \label{def sus}%
\end{equation}
therefore we have $F\simeq1-\chi\delta^{2}/2$.

In this paper, we study the RFS of two neighboring sites, and the RDM is
block-diagonal as $\rho=\varrho_{1}\oplus\varrho_{2}$ due to the specific
symmetries in the 1D TFIM, and $\varrho_{1}$, $\varrho_{2}$ are $2\times2$
matrices. In \cite{JianMa:arxiv/0805.4062v1}, we have derived general formulas
of RFS for density matrix of block-diagonal as $\rho=\oplus_{i=1}^{n}%
\varrho_{i}$, where $\varrho_{i}$'s are $2\times2$ matrices.
However, in this
work the density matrix satisfies $\det{\varrho_{i}}\neq0$ and $\text{tr}%
{\varrho_{i}}\neq0$ $\left(  i=1,2\right)  $, then we give the formula under
such case as follows~\cite{JianMa:arxiv/0805.4062v1},%
\begin{equation}
\chi_{i}=\frac{1}{4\text{tr}\varrho_{i}}\left\{  \left(  \text{tr}\varrho
_{i}^{\prime}\right)  ^{2}-4\det\varrho_{i}^{\prime}+\frac{\left[
\partial_{h}\det\left(  \varrho_{i}\right)  \right]  ^{2}}{\det\left(
\varrho_{i}\right)  }\right\}  , \label{general_fidelity}%
\end{equation}
where $\varrho^{\prime}\equiv\partial_{h}\varrho$, and $\chi_{i}$ is
the `susceptibility' for block $\varrho_{i}$. From the above formula
we can see that the RFS is only depends upon to the first
derivatives of the matrix elements.

\section{Scaling and Critical Behaviors of RFS in 1D TFIM}

The Hamiltonian of the 1D TFIM reads%

\begin{equation}
H_{I}=-\sum_{j=-M}^{M}\left[  \lambda\sigma_{j}^{x}\sigma_{j+1}^{x}+\sigma
_{j}^{z}\right]  , \label{Hamiltonian}%
\end{equation}
where $\sigma_{j}^{\alpha}$ $\left(  \alpha=x,y,z\right)  $ is a Pauli matrix
at site $j$, $\lambda$ is the Ising coupling constant in units of the
transverse field, and periodic boundary conditions $\left(  \sigma_{-M}%
=\sigma_{M}\right)$ are assumed. The total spins are $N=2M+1$, and
in this paper we consider $N$ is even, therefore $M$ is half odd
integer. As is well known, there is a critical point exactly at
$\lambda_{c}=1$ in the thermodynamic limit. The transition in this
model is of order to disorder type due to the competition between
the Ising coupling and the external magnetic field. For
$\lambda<\lambda_{c}$, the ground state of the system is
paramagnetic, while for $\lambda>\lambda_{c}$ the strong Ising
coupling introduces a magnetic long-range order of the order
parameter $\langle \sigma^{x}\rangle$ to the ground state.

The RDM of two neighboring sites takes the form of~\cite{XWang:EPJD/18}%

\begin{equation}
\rho_{i,i+1}=%
\begin{pmatrix}
u_{+} & z_{-} & 0 & 0\\
z_{-} & u_{-} & 0 & 0\\
0 & 0 & w & z_{+}\\
0 & 0 & z_{+} & w
\end{pmatrix}
, \label{2body_rdm}%
\end{equation}
in the basis $\left\{ |\uparrow\uparrow\rangle,|\downarrow\downarrow
\rangle,|\uparrow\downarrow\rangle,|\downarrow\uparrow\rangle\right\}
$ with $|\uparrow\rangle$ and $|\downarrow\rangle$ spin up and down.
The matrix
elements are%
\begin{align}
u_{\pm}  &  =\frac{1}{4}\left(  1\pm2\langle\sigma^{z}\rangle+\langle
\sigma_{0}^{z}\sigma_{1}^{z}\rangle\right)  ,\nonumber\\
w  &  =\frac{1}{4}\left(  1-\langle\sigma_{0}^{z}\sigma_{1}^{z}\rangle\right)
,\nonumber\\
z_{\pm}  &  =\frac{1}{4}\left(  \langle\sigma_{0}^{x}\sigma_{1}^{x}\rangle
\pm\langle\sigma_{0}^{y}\sigma_{1}^{y}\rangle\right)  . \label{mat_elem}%
\end{align}
where the translation invariance of the 1D TFIM is considered, and
we set $i=0$ for convenience in what follows. From Eqs.
(\ref{general_fidelity}) and (\ref{2body_rdm}), the RFS can be written explicitly with these matrix elements as
\begin{widetext}%
\begin{align}
\chi= &  \frac{1}{4\left(  u_{+}+u_{-}\right)  }\left[  \left(  \partial
_{\lambda}u_{+}-\partial_{\lambda}u_{-}\right)  ^{2}+4\left(  \partial
_{\lambda}z_{-}\right)  ^{2}+\frac{\left(  u_{-}\partial_{\lambda}u_{+}%
+u_{+}\partial_{\lambda}u_{-}-2z_{-}\partial_{\lambda}z_{-}\right)  ^{2}%
}{u_{+}u_{-}-z_{-}^{2}}\right]  \nonumber\\
&  +\frac{1}{2w}\left[  \left(  \partial_{\lambda}z_{+}\right)  ^{2}%
+\frac{\left(  w\partial_{\lambda}w-z_{+}\partial_{\lambda}z_{+}\right)  ^{2}%
}{w^{2}-z_{+}^{2}}\right]  .\label{RFS}%
\end{align}
\end{widetext}
The following discussions of critical behaviors of the RFS is based
on this equation.

The Hamiltonian could be diagonalized by using Jordan-Wigner, Fourier and
Bogoliubov transformation sequently (see \cite{SachdevQPTs}), and the mean
magnetization in the ground state is given by \cite{Barouch:PRA1075}%

\begin{equation}
\langle\sigma^{z}\rangle=\frac{1}{N}\sum_{q=-M}^{M}\frac{1-\lambda\cos\phi
_{q}}{\omega_{\phi_{q}}}, \label{sz}%
\end{equation}
where $\omega_{\phi_{q}}$ is the dispersion relation,%

\begin{align}
\omega_{\phi_{q}}  &  =\sqrt{1+\lambda^{2}-2\lambda\cos\phi_{q}},\nonumber\\
\phi_{q}  &  =2\pi q/N,
\end{align}
with $q$ half odd integer. The neighboring two-point correlation
functions are calculated as
\cite{Barouch:PRA786}%

\begin{align}
\langle\sigma_{0}^{x}\sigma_{1}^{x}\rangle &  =\frac{1}{N}\sum_{q=-M}^{M}%
\frac{\lambda-\cos\phi_{q}}{\omega_{\phi_{q}}},\nonumber\\
\langle\sigma_{0}^{y}\sigma_{1}^{y}\rangle &  =\frac{1}{N}\sum_{q=-M}^{M}%
\frac{\lambda\cos\left(  2\phi_{q}\right)  -\cos\phi_{q}}{\omega_{\phi_{q}}%
},\nonumber\\
\langle\sigma_{0}^{z}\sigma_{1}^{z}\rangle &  =\langle\sigma^{z}\rangle
^{2}-\langle\sigma_{0}^{x}\sigma_{1}^{x}\rangle\langle\sigma_{0}^{y}\sigma
_{1}^{y}\rangle. \label{correlation_func}%
\end{align}

It's known that the most important themes in critical phenomena are
scaling and universality. For a finite system, there are no
singularities unless the ground state level crossing occurs. In the
following, we will calculate the mean magnetization, the correlation
functions and their first derivatives, and get the finite-size
scaling and critical behaviors of RFS.

\subsection{Finite-size scaling behavior of RFS}

\begin{figure}
[ptb]
\begin{center}
\includegraphics[
height=2.3884in,
width=2.8496in
]%
{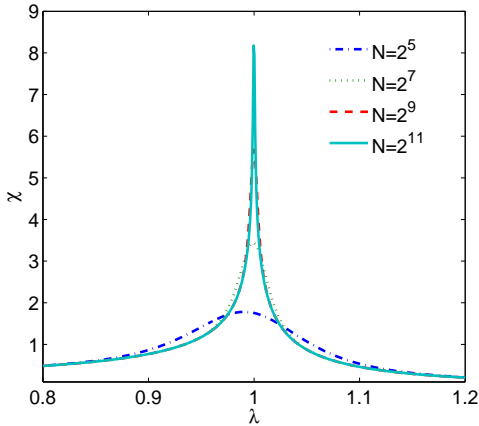}%
\caption{RFS as a function of $\lambda$ for various system sizes. The
positions of the peaks approach to the critical point at $\lambda=1$ as
increase with system sizes.}%
\label{sus_lambda}%
\end{center}
\end{figure}

The numerical results of the RFS for a finite-size system are shown
in Fig.~\ref{sus_lambda}, from which we can see that the peaks
become sharper and sharper as the system size increases, and their
locations approach to the critical point at the same time. It's
expected that there will be singular at $\lambda_{c}$ if $N$ is
infinite. Then to study the scaling behavior of RFS, we define its
maximum over $\lambda$ as $\chi_{m}$ with the location
$\lambda_{m}$.%
For a finite-size system, $\lambda_{m}\neq\lambda_{c}$, and yet if
$N$ is large enough, $\lambda_{m}$ is so close to $\lambda_{c}$ that
it's a good approximation for us to use $\lambda_{m}=\lambda_{c}$ in
computing $\chi_{m}$. The summations in Eqs.~(\ref{sz}) and
(\ref{correlation_func}) converge quickly as $N$ increases, which
means for large $N$, the summation can be
replaced by the integral as%
\begin{equation}
\lim_{N\rightarrow\infty}\frac{1}{N}\sum_{q=-M}^{M}\longrightarrow\frac
{1}{2\pi}\int_{-\pi}^{\pi}d\phi. \label{sum_integral}%
\end{equation}
Then it's easy to evaluate the integrals and get%
\begin{align}
\langle\sigma^{z}\rangle|_{\lambda=1}  &  =\frac{2}{\pi}, & \langle\sigma
_{0}^{x}\sigma_{1}^{x}\rangle|_{\lambda=1}  &  =\frac{2}{\pi},\nonumber\\
\langle\sigma_{0}^{y}\sigma_{1}^{y}\rangle|_{\lambda=1}  &  =-\frac{2}{3\pi},
& \langle\sigma_{0}^{z}\sigma_{1}^{z}\rangle|_{\lambda=1}  &  =\frac{16}%
{3\pi^{2}}. \label{expect_val}%
\end{align}
However, the first derivatives of the above quantities are divergent
with respect to $N$ at $\lambda_{c}$, and we have%
\begin{align}
\frac{d\langle\sigma^{z}\rangle}{d\lambda}\bigg{|}_{\lambda=1}  &  =-\frac
{1}{2N}\sum_{q=-M}^{M}\left\vert \frac{\cos^{2}\left(  \phi_{q}/2\right)
}{\sin\left(  \phi_{q}/2\right)  }\right\vert \simeq-\frac{1}{\pi}\ln
N,\nonumber\\
\frac{d\langle\sigma_{0}^{x}\sigma_{1}^{x}\rangle}{d\lambda}\bigg{|}_{\lambda
=1}  &  =\frac{1}{2N}\sum_{q=-M}^{M}\left\vert \frac{\cos^{2}\left(  \phi
_{q}/2\right)  }{\sin\left(  \phi_{q}/2\right)  }\right\vert \simeq\frac
{1}{\pi}\ln N,\nonumber\\
\frac{d\langle\sigma_{0}^{y}\sigma_{1}^{y}\rangle}{d\lambda}\bigg{|}_{\lambda
=1}  &  =\frac{1}{4N}\sum_{q=-M}^{M}\frac{\cos\left(  2\phi_{q}\right)
+\cos\phi_{q}}{\left\vert \sin\left(  \phi_{q}/2\right)  \right\vert }%
\simeq\frac{1}{\pi}\ln N.
\end{align}
The main contribution to the above expressions in the large-$N$ limit arises
from the summation around the zero point of $\sin(\phi_{q}/2)$, since%
\begin{equation}
\frac{1}{N}\sum_{q\in S}\frac{1}{\left\vert \phi_{q}\right\vert }\simeq
\frac{1}{\pi}\ln N,
\end{equation}
where $S$ represents the singular area. Immediately, with the help
of Eqs.~(\ref{general_fidelity}) and (\ref{correlation_func}), we
get
\begin{equation}
\frac{d\langle\sigma_{0}^{z}\sigma_{1}^{z}\rangle}{d\lambda}\bigg{|}_{\lambda
=1}\simeq-\frac{16}{3\pi^{2}}\ln N.
\end{equation}
The matrix elements and their derivatives could be evaluated by
inserting the above results into Eq.~(\ref{mat_elem}). Here we
should specifically consider
$\partial_{\lambda}z_{-}=\partial_{\lambda}\left(  \langle\sigma_{0}^{x}%
\sigma_{1}^{x}\rangle-\langle\sigma_{0}^{y}\sigma_{1}^{y}\rangle\right)  /4$,
the divergent behaviors of $\partial_{\lambda}\langle\sigma_{0}^{x}\sigma
_{1}^{x}\rangle$ and $\partial_{\lambda}\langle\sigma_{0}^{y}\sigma_{1}%
^{y}\rangle$ at $\lambda_{c\text{ }}$are both $(\ln N)/\pi$, while their
difference is convergent and could be integrated as a constant:%
\begin{equation}
\frac{dz_{-}}{d\lambda}\bigg{|}_{\lambda=1}=\frac{1}{3\pi}.
\end{equation}
Then, with Eq.~(\ref{RFS}), we get the divergent form of $\chi_{m}$
with respect to $N$ as
\begin{equation}
\chi_{m}\left(  N\right)  \simeq A_{1}\left(  \ln N+c_{1}\right)  ^{2}+c_{2},
\end{equation}
where $c_{1}$, $c_{2}$ are constants that could not be determined
analytically, and the coefficient%
\begin{align}
A_{1}  &  =\frac{27\pi^{4}-144\pi^{2}-1024}{\pi^{2}\left(
9\pi^{2}+32\right)\nonumber
\left(  3\pi^{2}-32\right)  +4096}\\
&  \simeq0.1485.
\end{align}
To perform a finite-size scaling analysis, we should find the
logarithmic divergent quantity, and the above result suggests us to
take the square root
of $\chi_{m}$ for large $N$ to get the logarithmic divergent form as%
\begin{equation}
\chi_{m}\left(  N\right)  ^{1/2}\simeq\sqrt{A_{1}}\ln N+\text{const.}%
\end{equation}%
\begin{figure}
[ptb]
\begin{center}
\includegraphics[
trim=0.000000in 0.000000in -0.261625in 0.000000in,
height=2.4118in,
width=2.8501in
]%
{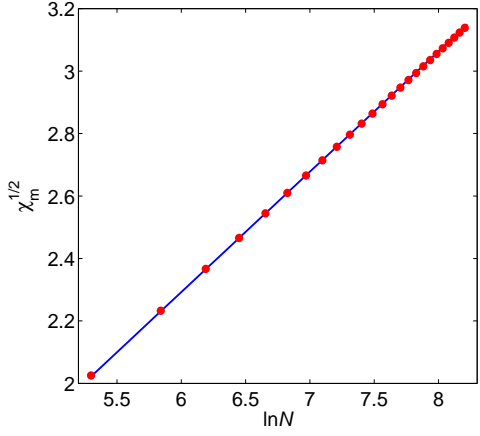}%
\caption{Square of the maximum RFS over $\lambda$ as a function of $N$. The
red dots are numerical results and the slope of the solid line is $\sqrt
{A_{1}}$.}%
\label{sus_log_n}%
\end{center}
\end{figure}
To confirm this, we compute $\chi_{m}^{1/2}$ numerically, and the
result is shown in Fig.~\ref{sus_log_n}, from which we can see that
the slope is excellently consistent with $\sqrt{A_{1}}$, while the
linear relation is obvious for not large $N$ because $c_{2}$ is
small with numerical computation. However the maximum global ground
state fidelity susceptibility $S_{m}$ is of a different scaling form
as $\ln S_{m}=2\ln N+$const~\cite{Chen:arxiv/0801.0020}, thus $\ln
S_{m}$ but not $S_{m}$ will exhibit similar scaling behavior with
$\chi_{m}^{1/2}$ at the critical point.

\subsection{Critical behavior of RFS in the thermodynamic limit}

In the thermodynamic limit the total spin $N$ is infinite, then the
summations can be replaced by the integrals (\ref{sum_integral}) and
are evaluated as:
\begin{align}
\langle\sigma^{z}\rangle= &  \frac{1-\lambda}{\pi}\text{K}\left(  k\right)
+\frac{1+\lambda}{\pi}\text{E}\left(  k\right)  ,\nonumber\\
\langle\sigma_{0}^{x}\sigma_{1}^{x}\rangle= &  \frac{\lambda-1}{\pi}%
\text{K}\left(  k\right)  +\frac{1+\lambda}{\pi}\text{E}\left(  k\right)
,\nonumber\\
\langle\sigma_{0}^{y}\sigma_{1}^{y}\rangle= &  \frac{1}{3\pi\lambda
}\big{[}\text{K}\left(  k\right)  \left(  \lambda-1\right)  \left(
2\lambda^{2}+1\right)  \nonumber\\
&  -\text{E}\left(  k\right)  \left(  \lambda+1\right)  \left(  2\lambda
^{2}-1\right)  \big{]},\label{elliptic}%
\end{align}
where K$\left(  k\right)  $ is the complete elliptic integral of the
first kind and E$\left(  k\right)  $ is the elliptic integral of the
second kind with $k=2\sqrt{\lambda}/\left(  1+\lambda\right)  $.
Therefore the critical behavior of the RFS is determined by the
elliptical integrals. At $\lambda_{c}$, we have
$k_{c}=2\sqrt{\lambda_{c}}/\left(  1+\lambda _{c}\right)  =1$, and
E$\left(  1\right)  =1$, while K$\left(  k\right)  \,$ is divergent
as $k\rightarrow1$, its asymptotic behavior is
\begin{equation}
\text{K}\left(  k\right)  \simeq\ln\frac{1}{\sqrt{1-k^{2}}}=\ln\frac
{1}{\left\vert 1-\lambda\right\vert },\label{EllipticK}%
\end{equation}
from which we have K$\left(  k_{c}\right)  \left(  1-\lambda_{c}\right)  =0$.
Then, obviously, at the critical point, with the above analysis we get the
same results with (\ref{expect_val}) in the previous subsection.

However, their derivatives are singular at $\lambda=1$. We take $\langle
\sigma^{z}\rangle$ for example,
\begin{equation}
\frac{d\langle\sigma^{z}\rangle}{d\lambda}=\frac{\lambda+1}{\pi\lambda
}\text{E}\left(  k\right)  -\frac{\lambda^{2}+1}{\pi\lambda\left(
\lambda+1\right)  }\text{K}\left(  k\right)  ,
\end{equation}
where we have used the following relations
\begin{align}
\frac{d\text{K}\left(  k\right)  }{dk} &  =\frac{\text{E}\left(  k\right)
}{k}-\frac{\text{K}\left(  k\right)  }{k},\nonumber\\
\frac{d\text{E}\left(  k\right)  }{dk} &  =\frac{\text{E}\left(  k\right)
}{\left(  1-k^{2}\right)  k}-\frac{\text{K}\left(  k\right)  }{k}.
\end{align}
Then as $\lambda$ approaches to $\lambda_{c}$, E$(k)$ converges quickly to 1,
and with Eq.~(\ref{EllipticK}) we get%
\begin{equation}
\frac{d\langle\sigma^{z}\rangle}{d\lambda}\simeq-\frac{1}{\pi}\ln\frac
{1}{\left\vert 1-\lambda\right\vert },
\end{equation}
which is logarithmic divergent as $\lambda\rightarrow\lambda_{c}$.%
\begin{figure}
[ptb]
\begin{center}
\includegraphics[
height=2.3632in,
width=2.9469in
]%
{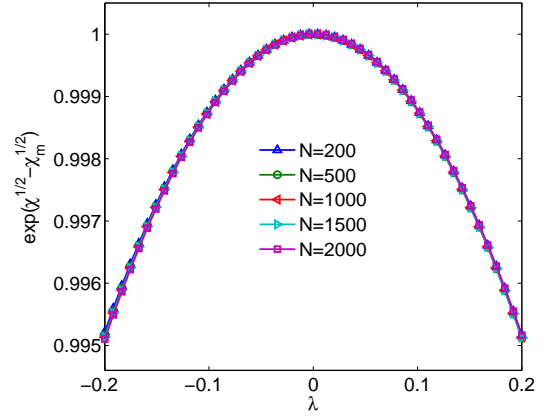}%
\caption{Finite-size scaling analysis is performed. The RFS as a
function of system size and coupling, collapses on a single curve
for various system
sizes.}%
\label{scabeh_sus}%
\end{center}
\end{figure}
The derivatives of the rest correlation functions are
\begin{align}
\frac{d\langle\sigma_{0}^{x}\sigma_{1}^{x}\rangle}{d\lambda} &  \simeq\frac
{1}{\pi}\ln\frac{1}{\left\vert 1-\lambda\right\vert },\nonumber\\
\frac{d\langle\sigma_{0}^{y}\sigma_{1}^{y}\rangle}{d\lambda} &  \simeq\frac
{1}{\pi}\ln\frac{1}{\left\vert 1-\lambda\right\vert },\nonumber\\
\frac{d\langle\sigma_{0}^{z}\sigma_{1}^{z}\rangle}{d\lambda} &  \simeq
-\frac{16}{3\pi^{2}}\ln\frac{1}{\left\vert 1-\lambda\right\vert },
\end{align}
where the logarithmic divergent terms come from the elliptic
integral K$\left(  k\right)  $. It's noticed that the coefficients
of the logarithmic terms are the same with those in the previous
subsection for a finite-size system.
Then we get a similar divergent form of $\chi$ as%
\[
\chi\left(  \lambda\right)  \simeq A_{2}\left(  \ln\frac{1}{\left\vert
1-\lambda\right\vert }+d_{1}\right)  ^{2}+d_{2},
\]
where $A_{2}=A_{1}$, and $d_{1}$, $d_{2}$ are constants that could not be
determined analytically.

As we have derived in the above, the divergent form of the RFS
around the critical point is $\left(  \ln N\right)  ^{2}$ for finite
system and $\left( \ln\left\vert 1-\lambda\right\vert \right)  ^{2}$
in the thermodynamic limit. In order to make a finite size scaling
analysis, it's feasible for us to use the square root of the RFS,
which is logarithmic divergent, and according to the scaling ansatz
\cite{BarberPhaseTran}, the square root of RFS is a function of
$N^{\nu}\left(  \lambda-\lambda_{m}\right)  $, which behaves as
$\chi\left(  \lambda_{m},N\right)  ^{1/2}-\chi\left(
\lambda,N\right) ^{1/2}\sim Q[N^{\nu}\left(  h-h_{m}\right)  ]$,
where $Q\left(  x\right) \propto\ln\left(  x\right)  $ for large
$x$. This function is universal and does not depend on system size
$N$. Since $A_{2}=A_{1}$, we determine the scaling exponent $\nu=1$
which is the same with the concurrence \cite{Osterloh:Nature416} and
the global ground state fidelity susceptibility
\cite{Chen:arxiv/0801.0020}, and is confirmed numerically shown in
Fig.~\ref{scabeh_sus}.

\section{Conclusion}

In summary, we have studied the scaling and critical behaviors of
the RFS for two neighboring sites in the 1D TFIM. We found that the
divergent behaviors of the RFS take the form of $\left( \ln N\right)
^{2}$ for finite-size system and $\left( \ln\left\vert
\lambda-1\right\vert \right)  ^{2}$ in the thermodynamic limit,
which are distinct from other quantities like the global ground
state fidelity susceptibility \cite{Chen:arxiv/0801.0020},
concurrence \cite{Osterloh:Nature416} and entanglement entropy
\cite{S-J-Gu:PRA032308}. Then we use the square root of the RFS
which is logarithmic divergent to carry out a finite-size scaling
analysis. The scaling exponent is determined analytically, and is
confirmed numerically. It's shown that, the RFS undergoes
singularity around the critical point, thus indicate that the RFS
can be used to characterize the QPTs.

\section{Acknowledgements}

We are indebted to Shi-Jian Gu, C. P. Sun and Z. W. Zhou for fruitful and
valuable discussions. The work was supported by the Program for New Century
Excellent Talents in University (NCET), the NSFC with grant No.~90503003, the
State Key Program for Basic Research of China with grant No.~2006CB921206, the
Specialized Research Fund for the Doctoral Program of Higher Education with
grant No.~20050335087.

\end{document}